\pdfoutput=1
\documentclass[12pt,aps,onecolumn,notitlepage,%
nofootinbib,preprintnumbers,superscriptaddress,longbibliography,%
amsmath,amssymb]{revtex4-2}
\usepackage{graphicx}
\usepackage{bm}
\usepackage[breaklinks,colorlinks=true]{hyperref}

\newcommand{\fm}{\,\text{fm}}

\newcommand{\GeVfm}{\,\text{GeV/fm$^3$}}
\newcommand{\Msun}{M_{\odot}}
\usepackage{ulem}

\begin{document}

\title{Gravitational Wave Signal for Quark Matter\\
  with Realistic Phase Transition}

\author{Yuki Fujimoto}
\affiliation{Department of Physics, The University of Tokyo,
  Tokyo 113-0033, JAPAN}
\affiliation{Institute for Nuclear Theory, University of Washington, Seattle, WA 98195, USA}

\author{Kenji Fukushima}

\affiliation{Department of Physics, The University of Tokyo,
  Tokyo 113-0033, JAPAN}

\author{Kenta Hotokezaka}

\affiliation{Research Center for the Early Universe (RESCEU),
  The University of Tokyo, Tokyo 113-0033, JAPAN}

\author{Koutarou Kyutoku}

\affiliation{Department of Physics, Kyoto University,
  Kyoto 606-8502, JAPAN}
\affiliation{Center for Gravitational Physics and Quantum Information,
  Yukawa Institute for Theoretical Physics, Kyoto University, Kyoto
  606-8502, JAPAN}
\affiliation{Interdisciplinary Theoretical and Mathematical Sciences
  Program (iTHEMS), RIKEN, Saitama 351-0198, JAPAN}

\begin{abstract}
  \vspace*{1em}\noindent
  {\large Abstract}\\
The cores of neutron stars (NSs) near the maximum mass realize the
most highly compressed matter in the universe where quark degrees of
freedom may be liberated.  Such a state of dense matter is
hypothesized as quark matter (QM) and its presence has awaited to be
confirmed for decades in nuclear physics.  Gravitational waves from
binary NS mergers are expected to convey useful information called the
equation of state (EOS).  However, the signature for QM with realistic
EOS is not yet established.  Here, we show that the gravitational wave
in the post-merger stage can distinguish the theory scenarios with and
without a transition to QM\@.  Instead of adopting specific EOSs as
studied previously, we compile reliable EOS constraints from the
\textit{ab initio} approaches.  We demonstrate that early collapse to
a black hole after NS merger signifies softening of the EOS associated
with the onset of QM in accord with \textit{ab initio} constraints.
Nature of hadron-quark phase transition can be further constrained by
the condition that electromagnetic counterparts need to be energized
by the material left outside the remnant black hole.
 \end{abstract}
\maketitle
\newpage



Heavy nuclei have a property called saturation and the baryon density
in central regions inside nuclei remains almost constant around
$n_0\sim 0.16\fm^{-3}$.  The corresponding rest-mass density is
$\rho_0 = m_N n_0 \sim 2.7\times 10^{14}\,\text{g/cm$^3$}$ with $m_N$
being the nucleon mass.  This is already a density of unimaginable
order, but in the universe we can find even denser matter compressed
with the help of gravitational force.  Among various compact stellar
objects neutron stars (NSs) are unique systems with complicated
structures having its typical mass around $\sim 1.4\Msun$, where
$\Msun$ denotes the solar mass, and its estimated size of
$\sim 10\,\text{km}$.  In the NS environment the gravitational force
is so strong that the saturation property of nuclear matter (NM) can
be overwhelmed and the maximum rest-mass density can reach
$\sim 5\rho_0$ or even higher.  The quantitative determination of the
maximum rest-mass density in the NS requires a relation of the
pressure in response to gravitational compression.  There have been a
huge number of theoretical and experimental attempts to extract
information about the equation of state (EOS); that is, the pressure
$p$ as a function of the rest-mass density $\rho$ of NS
matter~\cite{Lattimer:2015nhk,Ozel:2016oaf,Baym:2017whm}.

Since seminal works~\cite{Itoh:1970uw,Collins:1974ky} on the
hypothetical existence of quark matter (QM) in NSs, there have been
disputes on the interpretation of NS observational data.  For example,
it was argued~\cite{Ozel:2006bv} that EXO~0748-676 would rule out the
soft EOS and thus the presence of QM in the NS cores.  Soon later,
however, a counter argument~\cite{Alford:2006vz} appeared to claim
that EXO~0748-676 can be consistent with some EOS variants in the
presence of QM\@.  The discovery of a two-solar-mass NS
(PSR$\,$J1614-2230) in 2010~\cite{Demorest:2010bx} gave more decisive
impacts to the nuclear physics community.  The possible EOS candidates
are severely constrained and a strong first-order phase transition up
to a certain density has been excluded without
doubt~\cite{Alford:2015dpa}.  Now, there are similar radio
measurements that have confirmed massive NSs:
PSR$\,$J0348+0432~\cite{Antoniadis:2013pzd} and
PSR$\,$J0740+6620~\cite{NANOGrav:2019jur}.
A more comprehensive test of the EOS candidates with available data
including collective properties measured in the terrestrial experiment
of heavy-ion collision has also been proposed~\cite{Klahn:2006ir}.

In principle, the EOS is to be identified from the first-principles
theory of the Strong Interaction, i.e., quantum chromodynamics (QCD)
in terms of quarks and gluons~\cite{Ghiglieri:2020dpq}.  The leading
order (LO) in perturbative QCD (pQCD) expansion leads to thermodynamic
quantities of a non-interacting quark gas.  From the
next-to-next-to-LO (N$^2$LO), a logarithmic term involving the
renormalization scale emerges, and the EOS suffers theoretical
uncertainty in resumming logarithmic
singularities~\cite{Freedman:1976ub}.  The possible EOS window
including the strange quark mass effect has been carefully
quantified~\cite{Kurkela:2009gj}, and the resultant EOS can be
reliable at high enough density (see Ref.~\cite{Gorda:2021znl} for the
state-of-the-art pQCD EOS).  Alternative to QCD at low density in the
vicinity of $\sim\rho_0$ is the effective theory approach with pion
and nucleon degrees of freedom.  The chiral effective theory
($\chi$EFT) is based on the derivative expansion of QCD and it is
regarded as an \textit{ab initio} approach~\cite{Drischler:2021kxf}.
So far, the next-to-next-to-next-to-LO (N$^3$LO) calculation from the
$\chi$EFT has been applied for the nuclear
properties~\cite{Drischler:2017wtt} and the analysis of NS
matter~\cite{Drischler:2020fvz}.

In this way, the $\chi$EFT at low density and the pQCD at high density
tighten the favored region of the EOS
variations~\cite{Kurkela:2014vha} and the NS observation can further
constrain possible parameters.  The recent
analysis~\cite{Annala:2019puf} suggests EOS softening at a density
around $(5$--$6)\rho_0$, which may well be interpreted as the onset of
QM in the NS cores.  The evidence for QM is, however, not quite
conclusive yet and the multimessenger analysis is
indispensable~\cite{Alford:2019oge}.  Among various NS observations,
gravitational waves from binary NS mergers would be a promising probe
into the state of dense matter beyond the presumable QM onset.

In 2017 the LIGO-Virgo collaboration reported the first observation of
the gravitational wave from the binary NS merger in
GW170817~\cite{LIGOScientific:2017vwq}, which was two years later from
the first detection of the merger of the binary black holes (BHs) by
LIGO\@.  For GW170817, the gravitational wave from the inspiral stage
was identified, and the total mass of the system is found to be
$2.74^{+0.04}_{-0.01}M_\odot$ (90\% credibility).  The tidal
deformability was extracted, and the EOS in the intermediate density
region has been constrained
accordingly~\cite{Annala:2017llu,LIGOScientific:2018cki}.  In the
future, it is expected that third-generation detectors can find
several tens of binary NS mergers per year.  They will also enable us
to achieve high-signal-to-noise-ratio detection, so that the
uncertainty window should be narrowed and the information on the phase
transition may eventually be extracted~\cite{Raaijmakers:2021uju}.
While it is more challenging to detect the post-merger signals, they
are more sensitive to the phase transition to QM\@.  It is, therefore,
of utmost importance to make theoretical predictions for the
gravitational wave signals at the post-merger stage to seek for a hint
of QM\@.  Along these lines, numerical simulations have found that
significant effects on the gravitational wave should result from an
assumed strong first-order phase transition to
QM~\cite{Bauswein:2018bma,Most:2018eaw,Weih:2019xvw}.  The
multimessenger signals associated with supernova explosion and NS
mergers are also discussed~\cite{Bauswein:2022vtq}.

At the same time, a continuous transition (called ``crossover'' in the
QCD context) toward QM is theoretically supported.  In the presence of
color superconducting states, NM and QM can have identical global
symmetry, implying duality between the confining phase with baryons
and mesons and the Higgs phase with quarks and
gluons~\cite{Schafer:1998ef} (see also
Refs.~\cite{Cherman:2018jir,Hirono:2018fjr} for discussions on
topological phase transition).  This idea of continuity can be
generalized to a more realistic situation with finite strange quark
mass~\cite{Fujimoto:2019sxg}.  Moreover, it has been known that NM and
QM cannot be distinguished in an idealized world with the infinitely
large number of gluon species (i.e., large-$N_{\text{c}}$ limit), and
this dual nature of dense matter is referred to as Quarkyonic
Matter~\cite{McLerran:2007qj}.  It is notable that Quarkyonic Matter
can give a natural account~\cite{McLerran:2018hbz} for a possible peak
in the speed of sound with increasing density as seen from the NS data
inference~\cite{Fujimoto:2019hxv,Fujimoto:2021zas}.  From this point
of view, the merger simulations with preferable EOS candidates with
continuous crossover to QM should be important.  Recently, numerical
simulations with quark-hadron crossover (QHC) EOSs have been
reported~\cite{Huang:2022mqp,Kedia:2022nns}.


\begin{figure}
  \includegraphics[width=0.65\textwidth]{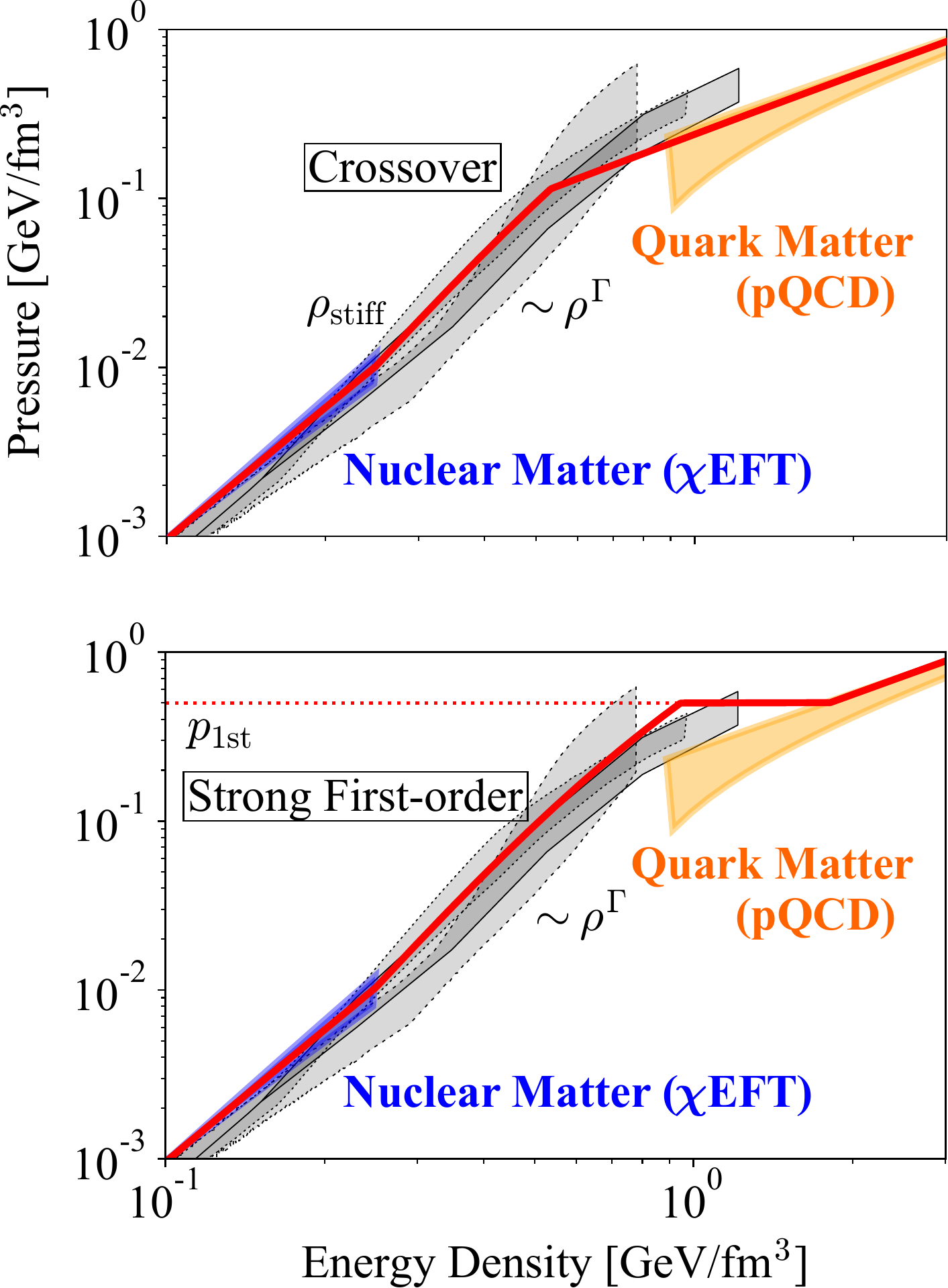}
  \caption{EOS candidates with a crossover (top) and a strong
    first-order transition (bottom) interpolated between $\chi$EFT
    (nuclear branch) and pQCD (quark branch).  The blue band
    represents the $1\sigma$ uncertainty in $\chi$EFT and the orange
    band represents the renormalization scale uncertainty from $X=2$
    (lower) to $X=4$ (upper) using the notation in
    Ref.~\cite{Fraga:2013qra}.  Gray bands show the data-driven EOSs
    inferred from the Bayesian (edged by
    dashed~\cite{Raaijmakers:2021uju} and dotted
    lines~\cite{Ozel:2015fia,Bogdanov:2016nle}) and deep learning
    (edged by a solid line~\cite{Fujimoto:2019hxv,Fujimoto:2021zas})
    analyses.}
  \label{fig:eos}
\end{figure}

We would remind that the \textit{ab initio} constraints have not been
properly taken into account for any gravitational wave simulations
(see Ref.~\cite{Gorda:2022jvk} for the latest analysis of the
\textit{ab initio} constraints).  Because QCD is the most fundamental
theory, if the density is high enough to justify the perturbative
expansion, the genuine EOS must eventually meet the quark branch from
pQCD\@.  If the phase transition onto the quark branch takes place at
higher density, a discontinuous jump is inevitable, leading to a
stronger first-order transition.  For lower transition densities the
observational data can constrain the EOS better.  In this way we can
classify theory possibilities into several distinct scenarios
according to the critical density of the phase transition.
Figure~\ref{fig:eos} summarizes two representative scenarios.

Now, we shall explain the common ingredients for the EOS
construction.  First of all, we note that the terms, ``soft'' and
``stiff'', refer to the EOS with $p$ relatively low and high,
respectively, for a given density, $\rho$, or the energy density,
$\varepsilon$.  At low density
($\varepsilon \lesssim 0.25\,\GeVfm$ in Figure~\ref{fig:eos}), a blue
shaded region labeled as ``Nuclear Matter'' represents the nuclear
branch, that is, a theoretical window predicted from N$^3$LO
$\chi$EFT~\cite{Drischler:2020fvz}.  At high density
($\varepsilon \gtrsim 1\,\GeVfm$ in Figure~\ref{fig:eos}), on the
other hand, the pQCD prediction is shown by an orange shaded region
(quark branch) labeled as ``Quark Matter'' on the figure.  Although
these are reliable constraints on the genuine EOS, we still need to
introduce (at least) two parameters corresponding to the validity
limits of two branches.  Here, one parameter, $\rho_{\text{stiff}}$,
represents the upper limit of the $\chi$EFT prediction, above which
the EOS must become stiff to support massive NSs.  The unknown
intermediate region between the nuclear and the quark branches may be
parametrized by piecewise polytrope: for our purpose of demonstrating
the distinguishability of crossover transition the intermediate EOS is
represented by a single polytrope, $p(\rho) = K\, \rho^\Gamma$ with
the adiabatic index, $\Gamma$.  Detailed shapes will be constrained by
the future NS observations~\cite{Forbes:2019xaz, Landry:2020vaw}.
Now, let us look into representative scenarios in theory.
\vspace{1em}


\noindent 
\textit{Crossover}

Strictly speaking, the category of smooth crossover would not exclude
a second-order nor a weak first-order phase transition.  The EOS is
continued to QM at the density when the intermediate EOS overshoots
the quark branch.  The quark branch has uncertainty from
renormalization scale, but the stiffest edge of the uncertainty band
is the most favored;  The smoothly connected EOS over the quark branch
should support massive NSs and it is very hard to satisfy this
condition unless the stiffest one is adopted.  This choice is also
consistent with the more convergent EOS from the resummed perturbation
theory~\cite{Fujimoto:2020tjc}.  We have performed extensive analyses
of $\rho_{\text{stiff}}$ and $\Gamma$ allowed by two conditions of the
causality and the two-solar-mass bound (see Method).  For a realistic
set of parameters, we have chosen $\rho_{\text{stiff}}=1.6\rho_0$ and
$\Gamma=3.5$ to draw the solid line in the top panel of
Figure~\ref{fig:eos}.  It is noteworthy that our choice of the
intermediate EOS looks consistent with observational data-driven EOSs
based on the Bayesian
analysis~\cite{Steiner:2010fz,Steiner:2012xt,Ozel:2015fia,Bogdanov:2016nle,Raaijmakers:2021uju}
as well as the deep
learning~\cite{Fujimoto:2019hxv,Fujimoto:2021zas}.
\vspace{1em}

\noindent 
\textit{Strong First-order Phase Transition at High Density}

In the first-order phase transition case, one more parameter,
$p_{\text{1st}}$, is necessary to specify the critical pressure.  The
bottom panel in Figure~\ref{fig:eos} shows an EOS with the strong
first-order phase transition for $p_{\text{1st}}=0.5\GeVfm$ with
$\rho_{\text{stiff}}=1.6\rho_0$ and $\Gamma=3.5$ unchanged.  We have
numerically confirmed that the NS structure is insensitive to the
phase transition up to the maximum-mass configuration for
$p_{\text{1st}}\gtrsim 0.4\GeVfm$.  This is because
$p_{\text{1st}} \gtrsim 0.4\GeVfm$ is higher than the pressure
achievable in the NS cores.  The EOS on the quark branch is softer
than most of the empirical EOSs that are phenomenologically accepted,
and a strong first-order phase transition is unavoidable in order to
match the quark branch.  Such a scenario of the strong first-order
phase transition is logically possible, though it is unlikely that
supposedly valid pQCD could describe two competing vacua there.
\vspace{1em}

As quantified in Method, other parameter sets are still possible, but
we must first confirm detectable differences for the representative
cases with and without crossover;  only after this confirmation we are
demonstrating in the present work, more systematic survey would make
sense.  Moreover, we note that a first-order transition in the
intermediate density region is disfavored; see Method for more details
and also discussions in Ref.~\cite{Komoltsev:2021jzg}.  Here, our
terminology, ``without crossover'', specifically means a possibility
of stiff EOSs (as the most empirical EOSs) followed by a strong
first-order phase transition at too high density to affect the
gravitational wave signal.  Then, for this comparison $p_{\text{1st}}$
is no longer a relevant parameter.  For actual simulations for the
case without crossover, we implicitly assume a large $p_{\text{1st}}$
and used the extrapolated stiff EOS with slight modification not to
violate the causality;  $\Gamma=3.5\to 2.9$ in the high density region
corresponding to the quark branch in the crossover scenario.
\vspace{1em}


\begin{figure}
  \includegraphics[width=\textwidth]{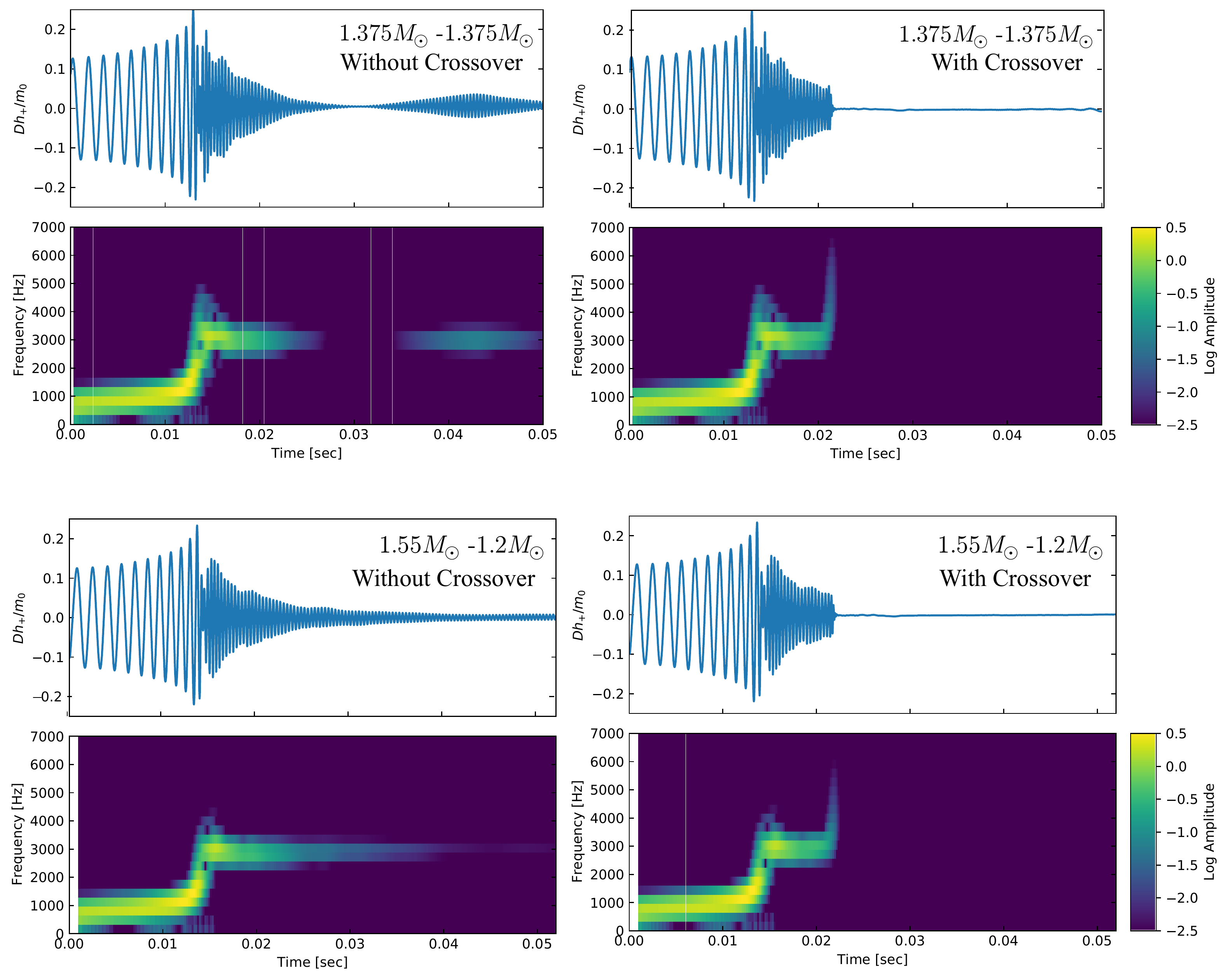}
  \caption{Plus-mode gravitational waveforms $h_+$ with the EOSs
    without crossover (left) and with crossover (right).  The signals
    are normalized as dimensionless $D h_+/m_0$ by the luminosity
    distance to the source, $D$, divided by the total mass of the
    system, $m_0$, in unit of $G=c=1$.  The corresponding spectrograms
    from the short-time Fourier transform are displayed below.}
  \label{fig:gw}
\end{figure}

Now, we come to discussions of our central results in
Figure~\ref{fig:gw}, where a merger simulation is performed with a
numerical relativity code \texttt{SACRA}~\cite{Yamamoto:2008js}.  The
upper and lower panels show the gravitational waves from binary NS
mergers with equal masses, $1.375\Msun$-$1.375\Msun$, and unequal
masses, $1.55\Msun$-$1.2\Msun$, respectively, chosen to be consistent
with GW170817~\cite{LIGOScientific:2017vwq}.  The left panels in
Figure~\ref{fig:gw} show the gravitational wave expected with the EOS
without crossover, i.e., a conventional EOS without rapid softening at
detectable density, while the right panels are results in the
crossover scenario.  In both scenarios, a remnant massive NS is
transiently formed after merger.  Because the EOS in the intermediate
density is characterized by a large value of $\Gamma$, the remnant
exhibits a non-axisymmetric, ellipsoidal
structure~\cite{Shibata:2005ss}.  Thus, the maximum density of the
system increases only moderately right after merger, and the
gravitational collapse does not occur immediately irrespective of the
crossover.  When the density eventually grows up above the QM onset
due to the hydrodynamical angular momentum transport and
gravitational-wave emission, the EOS softening associated with the
quark branch makes the transient NS collapse into a BH at
$\approx 7$--$8\,\textrm{ms}$ after the first bounce of merging NSs in
our models.  The time scale to the gravitational collapse depends
primarily on the total mass of the binary, which can be extracted from
the inspiral waveform with reasonable accuracy as in the case of
GW170817~\cite{LIGOScientific:2017vwq}, but only weakly on the mass
ratio of the system, the strength of the thermal effect (characterized
by the thermal index $\Gamma_{\text{th}}$: see Method), and the grid
resolution.  By contrast, if the transition to QM does not set in, the
remnant collapses (if possible) only after the long time scale of
magnetic dipole radiation (see, e.g., Ref.~\cite{Hotokezaka:2013iia})
thanks to the large maximum mass of the NS\@.  Accordingly, in this
case, the BH formation cannot be identified by the gravitational-wave
signal.

In the lower panels of Figure~\ref{fig:gw}, the corresponding
spectrograms from the short-time Fourier transform are shown.  It is
evident that the high frequency components are enhanced when the
gravitational collapse occurs, which signifies EOS softening induced
by the QM onset.  From this comparison of the gravitational-wave
patterns we conclude that the future measurement of the lifetime of
the remnant NS after merger and the comparison to systematic
simulations can constrain the presence/location of the softening point
in the EOS (QM onset).\footnote{One may think that the softening could
  be caused by hadronic effects such as hyperons, but according to
  Hagedorn's picture of deconfinement, the crossover is approached by
  liberation of more and more massive degrees of freedom; see
  Ref.~\cite{Andronic:2009gj} and references therein.}

\begin{figure}
  \includegraphics[width=0.55\textwidth]{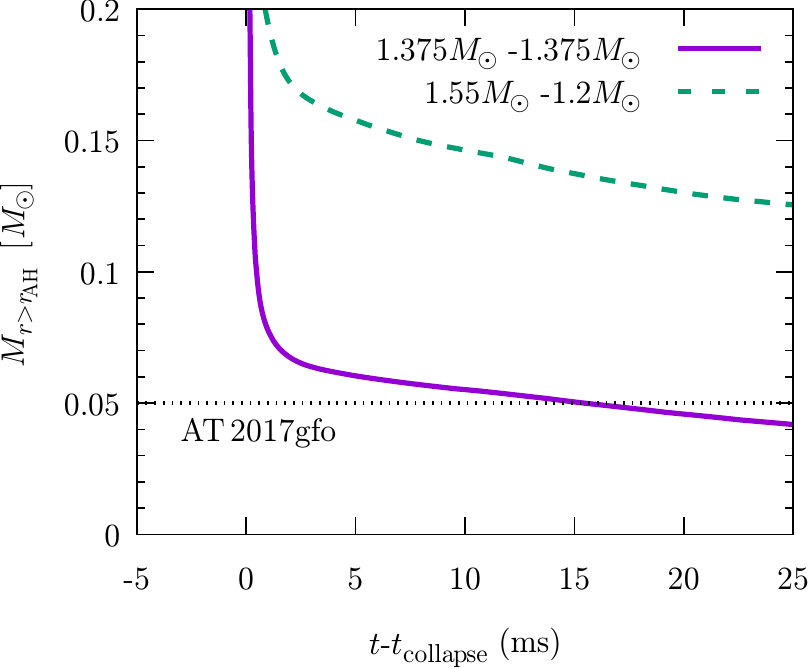}
  \caption{Remnant mass on the material outside the apparent horizon
    of the BH as a function of the time after the gravitational
    collapse, $t_{\text{collapse}}$.}
  \label{fig:disk}
\end{figure}

We finally emphasize that the crossover scenario is consistent with
multimessenger observations of GW170817. Specifically, it is widely
recognized that the associated kilonova, AT\;2017gfo, requires
ejection of $\approx 0.05M_\odot$ because of its
luminosity~\cite{Kasen:2018drm,Hotokezaka:2019uwo}.  For the total
mass of $2.75M_\odot$, such a substantial material can be ejected by
unequal-mass systems even if the crossover-induced collapse sets in.
Figure~\ref{fig:disk} shows the mass of material left outside the
apparent horizon (one variant of BH horizons),
$M_{r>r_\mathrm{AH}}$.  As the mass of the dynamical ejecta is
$<0.01M_\odot$ for the current models, $M_{r>r_\mathrm{AH}}$ needs to
be larger than $\approx 0.05M_\odot$ as a minimal requirement unless
the mass ejection is extremely efficient for
$\approx 7$--$8\,\textrm{ms}$ between the first bounce and the
gravitational collapse.  While the equal-mass binary violates this
criterion at $\approx 15\,\textrm{ms}$ after the gravitational
collapse, the unequal-mass binary may sustain $\gtrsim 0.05M_\odot$
much longer, giving greater potential for ejecting material
responsible for AT\;2017gfo. The precise mass and other
characteristics of ejected material may be clarified by
high-resolution simulations incorporating magnetic fields and
neutrinos (see, e.g., Ref.~\cite{Hayashi:2021oxy}), which we leave as
a topic for future studies along with calculations of nucleosynthesis
and electromagnetic emission.  In the coming future, the precise
measurements of binary masses with the gravitational-wave signals
together with the characterization with electromagnetic counterparts
for various mergers with different masses will enable us to constrain
the EOS at high density. Thus, multimessenger observations will
further delineate properties of QM and the nature of transition.

\bibliographystyle{utphys}
\bibliography{gw}

\section*{Acknowledgments}
We thank
Mark~Alford,
Aleski~Kurkela,
Sanjay~Reddy,
and
Wolfram~Weise
for discussions. 
This work was partially supported by
Japan Society for the Promotion of Science
(JSPS) KAKENHI Grant
No.\
20J10506 (YF),  
18H01211, 22H01216 (KF),
20K14513, 20H05639, 20H00158 (KH),
18H05236, 20H00158, 22K03617 (KK),
and JST FOREST Program Grant No.\ JPMJFR2136 (KH).

\section*{Author Contributions}
Y.F.\ made crucial suggestions on the EOS parametrization and checked
the EOS calculation consistency.
K.F.\ designed the EOS parametrization and performed the numerical
calculations of the EOSs.
K.H.\ gave discussions on the gravitational wave signals at the
inspiral stage and reformatted the spectrograms.
K.K.\ performed the numerical simulation for the gravitational wave
signals and analyzed the ejected and the remnant masses.
All authors equally contributed to writing the manuscript.

\section*{Author Information}
The authors declare no competing financial interests.
Interested readers can send comments to K.F.\
(fuku@nt.phys.s.u-tokyo.ac.jp) or Y.F.\
(yfuji@uw.edu) for the EOS construction and to
K.H.\ (kentah@g.ecc.u-tokyo.ac.jp) or K.K.\
(kyutoku@tap.scphys.kyoto-u.ac.jp) for the the simulation of numerical
relativity.

\renewcommand{\thefigure}{S\arabic{figure}}
\setcounter{figure}{0}

\section*{METHODS}

\noindent
{\underline{Piecewise Polytropic EOS:}}

The pressure in a piecewise density window $[\rho_{i-1}, \rho_i]$ is
given by
\begin{equation}
  p(\rho) = K_i\, \rho^{\Gamma_i}\,,
\end{equation}
and the continuity of the pressure to the adjacent segment requires,
\begin{equation}
  K_{i+1} = \frac{p(\rho_i)}{\rho_i^{\Gamma_{i+1}}}
  = K_i\, \rho_i^{\Gamma_i-\Gamma_{i+1}}\,.
\end{equation}
The energy density, $\varepsilon=-p+\mu n$, involves an integration
constant, $a_i$, and is written as
\begin{equation}
  \frac{\varepsilon}{\rho}
  = \frac{K_i}{\Gamma_i -1}\,\rho^{\Gamma_i-1} + 1 + a_i\,.
\end{equation}
Following the prescription in Ref.~\cite{Read:2008iy}, the integration
constant, $a_i$, is fixed by the condition to make $\varepsilon$
continuous, i.e.,
\begin{equation}
  a_{i+1} = a_i + \frac{\Gamma_{i+1} - \Gamma_i}
  {(\Gamma_{i+1}-1)(\Gamma_i-1)}\,K_i\,\rho_i^{\Gamma_i-1}\,.
\end{equation}
We note that the polytropic EOS is a function of $\rho$, not of the
chemical potential $\mu$, and this is why the integration constant is
needed;  see Ref.~\cite{Komoltsev:2021jzg} for related discussions.
\vspace{1em}

\noindent
{\underline{Constraining the Realistic EOS:}}

\begin{table}
\begin{tabular}{|c||c|c|c|}
    \hline
    $i$& $\rho_i$ & $\Gamma_i$ & $K_i$ \\
    \hline\hline
    Crust & $1.02148\mathrm{E}\!+\!14$ & $1.35692$ & $3.9987\mathrm{E}\!-\!8$ \\
    $\chi$EFT & $\rho_{\text{stiff}} = 4.28800\mathrm{E}\!+\!14$ & $2.64258$ & $3.50290\mathrm{E}\!-\!5$ \\
    Interpolated & $8.55829\mathrm{E}\!+\!14$ & $\Gamma=3.5$ & $9.96414\mathrm{E}\!-\!18$\\
    pQCD & --- & $1.45303$ & $3.66934\mathrm{E}\!+\!13$ \\
    \hline
  \end{tabular}
  \caption{Parameters in the polytropic EOS;  the pressure unit is
    $\mathrm{dyne}/\mathrm{cm}^2$ and the rest-mass density unit is
    $\mathrm{g}/\mathrm{cm}^3$.  The stiffening density is fixed as
    $\rho_{\text{stiff}}=1.6\rho_0=10^{14.632255}\,\mathrm{g}/\mathrm{cm}^3$.
    The crust parameters are adopted from Ref.~\cite{Read:2008iy}.}
  \label{tab:polytrope}
\end{table}

The EOS from N$^3$LO $\chi$EFT  can be well fitted by the polytropic
form with the parameters in Table~\ref{tab:polytrope}.  Actually the
uncertainty band allows for the soft and the stiff edges,
respectively, with $\Gamma\simeq 2.378$ and $2.832$, and we adopt the
middle value as a representative.  In the same way the pQCD EOS can
also be expressed by the polytropic form as listed in
Table~\ref{tab:polytrope}.  Once the nuclear branch at low density and
the quark branch at high density are determined, the EOS can be
uniquely specified with two remaining parameters:
$\rho_{\text{stiff}}$ and $\Gamma$ if the continuity is demanded.  For
the low-density region which corresponds to the NS crust and the
$\chi$EFT description is not valid, we adopt another single, soft
polytrope (see Table~\ref{tab:polytrope}).

\begin{figure}
  \includegraphics[width=0.62\textwidth]{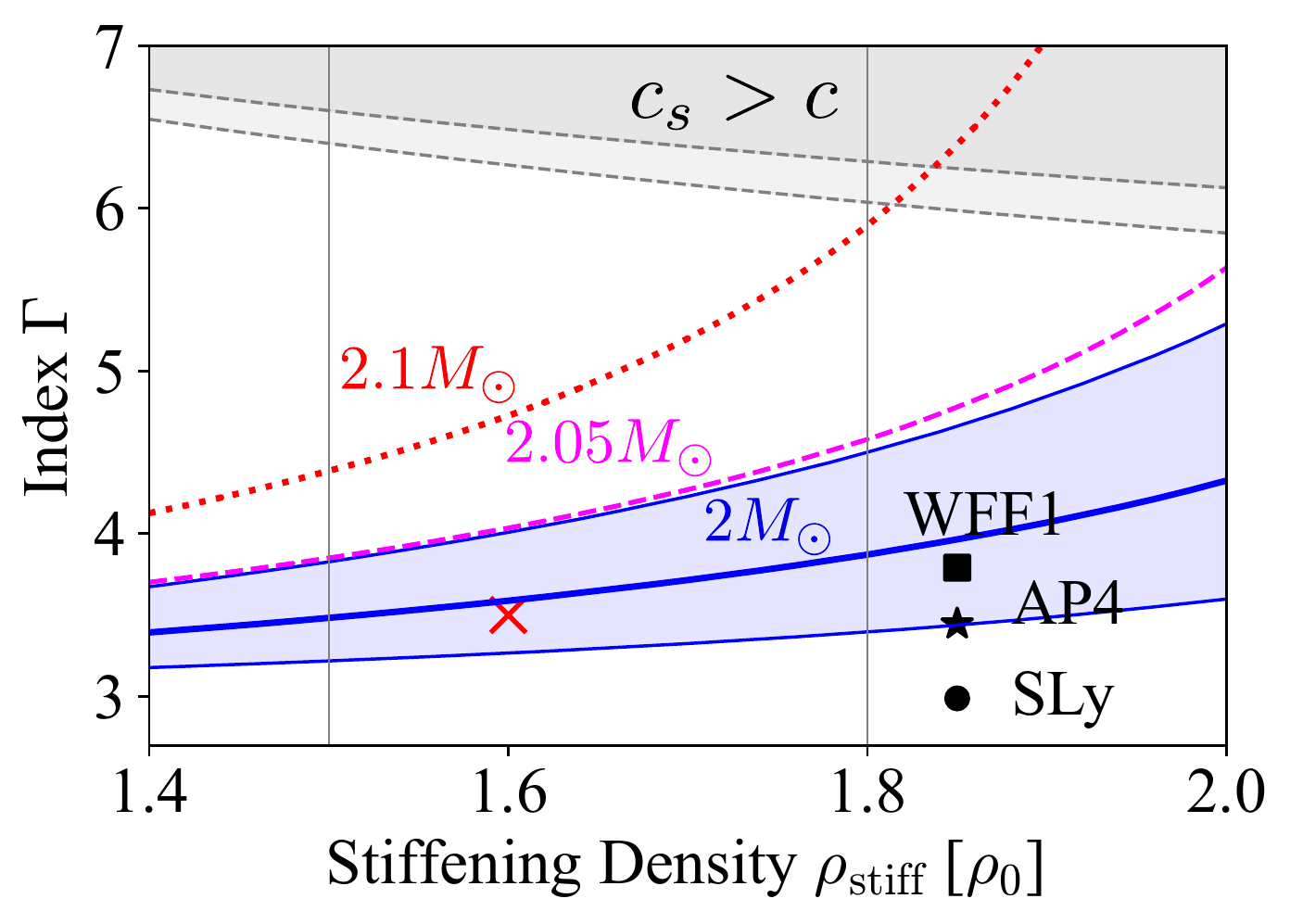}
  \caption{Allowed region of the stiffening density,
    $\rho_{\text{stiff}}$, and the adiabatic index, $\Gamma$.  The
    gray shaded region is excluded from the causality bound and the
    blue shaded region is the lower bound of $\Gamma$ to support the
    massive NS\@.  The red cross represent the parameters
    corresponding to our EOS choice.}
  \label{fig:region}
\end{figure}

Figure~\ref{fig:region} displays the allowed region of parameters.
The upper region (i.e., large-$\Gamma$ region) shaded with gray color
is excluded from the causality bound that the sound speed, $c_s$, must
not exceed the speed of light up to the maximum-mass configuration.
Two dashed lines represent the uncertainty band from the $\chi$EFT EOS
(and the quark branch is fixed at the stiffest edge).  The lower
region with small $\Gamma$ is bounded by the maximum mass of the NS,
which has a width associated with the uncertainty band from the
$\chi$EFT EOS  as well.  The blue shaded region represents the
condition that the maximum NS mass is $2M_{\odot}$.  The thick line in
the middle of the shaded region corresponds to our choice of the
parameters for the $\chi$EFT EOS and  the cross symbol to our choice
of $\rho_{\text{stiff}}$ and $\Gamma$ as given in
Table~\ref{tab:polytrope}.  The NS mass from our EOS is consistent
with pulsar observations within observational errors.  More systematic
studies on the choice of parameters are left for future studies.  If
the maximum mass is further raised to $2.05M_{\odot}$ and
$2.1M_{\odot}$, the EOS should be stiffer and $\Gamma$ should be
larger accordingly as shown by the purple dashed line and the red
dotted line, respectively.

In conventional EOSs, the reasonable values of $\Gamma$ ranges around
$2$--$4$.  For example, the comprehensive EOS list is found in
Ref.~\cite{Read:2008iy}, from which we took $\Gamma$'s of several
representative EOSs at the second density segment,
$\rho=(1.85$--$3.7)\rho_0$;
$\Gamma=3.791$ for WFF1, $\Gamma=3.445$ for AP4, and
$\Gamma=2.988$ for SLy are shown in Figure~\ref{fig:region} for
reference.  The present analysis aims to quantify visible impacts from
the EOS softening and we choose $\Gamma=3.5$ at
$\rho_{\text{stiff}}=1.6\rho_0$ for our demonstration purpose.

\begin{figure}
  \includegraphics[width=0.65\textwidth]{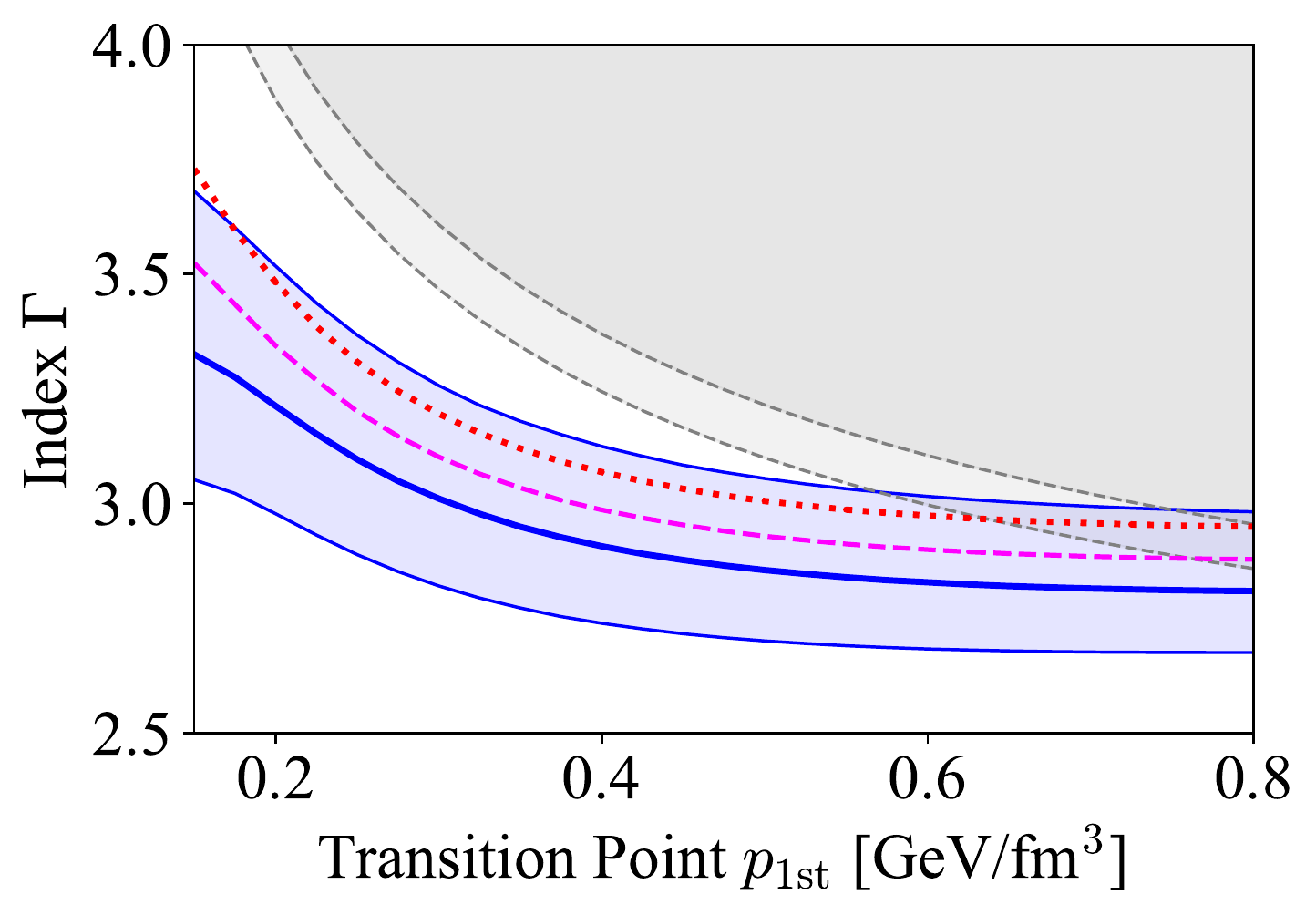}
  \caption{Allowed region of the first-order phase transition point,
    $p_{\text{1st}}$  and the adiabatic index, $\Gamma$.  The gray
    shaded region from the causality bound and the blue shaded region
    from the NS mass are shown in the same way as in
    Figure~\ref{fig:region}.}
  \label{fig:region_first}
\end{figure}

We can perform allowed parameter search in a similar fashion for the
EOS with a first-order phase transition.  In this case the first-order
phase transition point, $p_{\text{1st}}$, is involved in addition to
$\rho_{\text{stiff}}$ and $\Gamma$.  To draw
Figure~\ref{fig:region_first}, $\rho_{\text{stiff}}=1.6\rho_0$ is
chosen and the allowed region of $p_{\text{1st}}$ and $\Gamma$ is
plotted.  If $p_{\text{1st}}$ is higher than $\sim 0.4\GeVfm$, the
lower bound of $\Gamma$ is almost insensitive to $p_{\text{1st}}$
because the maximum pressure of the NS would no longer reach
$p_{\text{1st}}$.  Also, if the stiff EOS from the nuclear branch is
extended to the higher density, the causality bound is violated at
smaller $\Gamma$ and the allowed region becomes narrower as seen in
Figure~\ref{fig:region_first}.  Therefore, the allowed window is
already very limited.  The Bayesian analysis can estimate the
likelihood and the first-order phase transition is disfavored in the
detectable region~\cite{Komoltsev:2021jzg}.  If $p_{\text{1st}}$ is
large, in contrast, even the gravitational wave from the NS merger is
not influenced by phase transitions and the parameter search for
$p_{\text{1st}} \gtrsim 0.5\GeVfm$ is not crucial in practice.
\vspace{1em}

\noindent
{\underline{Thermal Index:}}

After the collision of NSs, shock interactions increase the
temperature to a few tens of MeV\@.  Because the cooling time scale
due to neutrino emission is as long as $\sim 1\,\mathrm{s}$, it is
necessary to incorporate thermal effects at least approximately.  The
thermal correction is included through
\begin{equation}
  p = p_{\text{cold}} + p_{\text{th}} \,,\qquad
  \varepsilon = \varepsilon_{\text{cold}} + \varepsilon_{\text{th}}\,,
\end{equation}
where the first term is the zero-temperature part and the second term
represents the thermal effect.  In numerical simulations, hydrodynamic
evolution equations determine the values of $\varepsilon$ and $\rho$.
Using the zero-temperature EOS, we may readily derive
$p_{\text{cold}}$ and $\varepsilon_{\text{cold}}$ from $\rho$.  Then,
the thermal index is introduced from an ideal-gas Ansatz:
\begin{equation}
  p_{\text{th}} = (\Gamma_{\text{th}}-1)\,\rho\,\varepsilon_{\text{th}}\,.
\end{equation}
The conventional choice of $\Gamma_{\text{th}}$ is within the range of
$1.5$--$1.8$, where a larger value typically gives a larger thermal
effect and a longer lifetime of the remnant NS\@.  The $\chi$EFT gives
a theory estimate of $\Gamma_{\text{th}}$ depending on the density.
According to Ref.~\cite{Carbone:2019pkr} a peak is located around
$\Gamma_{\text{th}}\sim 1.75$ around the density $\sim 0.5\rho_0$ and
$\Gamma_{\text{th}}$ decreases toward $\sim 1$ around the density
$\sim 2\rho_0$ (see also Ref.~\cite{Fujimoto:2021dvn} for model
calculations of $\Gamma_{\text{th}}(\rho)$).  In the present work the
gravitational collapse makes a sharp contrast to signify QM, and to
strengthen our argument, we choose the most conservative value of
$\Gamma_{\text{th}}=1.75$ for our simulation presented here.  This
choice is compatible with the preceding
studies~\cite{Bauswein:2018bma,Weih:2019xvw}.  We have numerically
checked that the qualitative behavior in the post-merger dynamics is
not changed for
$\Gamma_{\text{th}}=1.5$.\footnote{If $\Gamma_{\text{th}}=2$ is used,
  the thermal pressure can sustain the massive NS after the merger as
  presented in Ref.~\cite{Huang:2022mqp} and the BH formation may not
  be seen shortly after merger.  However, this choice of
  $\Gamma_{\text{th}}=2$ in Ref.~\cite{Huang:2022mqp} is motivated for
  a specific purpose to enhance the $f_2$ signal.}
In reality density-dependent $\Gamma_{\text{th}}$ could be even
smaller in some density range.  Besides, the effects of magnetic
fields and neutrinos are also likely to decrease the survival time of
the remnant after merger.
\vspace{1em}

\noindent
{\underline{Simulation:}}

Numerical simulations of binary NS mergers are performed with a
numerical-relativity code, \texttt{SACRA}~\cite{Yamamoto:2008js}.
This code has been well-tested and used for simulating compact object
binaries (see, e.g., Ref.~\cite{Hotokezaka:2013iia}).
Specifically, we solve the Einstein equation for gravity and ideal
hydrodynamics equations for matter.  Formulation adopted in this work
is described in Ref.~\cite{Kyutoku:2014yba}, in which the method for
constructing initial data is also presented.  Because we focus on the
short time scale of $\sim 20\,\mathrm{ms}$ after the first bounce,
magnetic fields or neutrino radiation transfer are not incorporated.

We mainly simulate two binary models with the total mass being fixed
to $2.75M_\odot$ motivated by GW170817~\cite{LIGOScientific:2017vwq}:
an equal-mass binary with $1.375M_\odot$--$1.375M_\odot$ and an
unequal-mass binary with $1.55M_\odot$--$1.2M_\odot$.  To check that
our numerical results are insensitive to the grid resolution, we
performed two simulations for each model. The finest grid spacing in
our adaptive-mesh-refinement structure is $\approx 190\,\textrm{m}$
and $\approx 280\,\textrm{m}$ for high- and low-resolution runs,
respectively. In all the models, the outer boundary is located at
$\approx 3000\,\mathrm{km}$ from the center of mass with the aid of
the adaptive-mesh-refinement algorithm. Gravitational waves are
extracted at $\approx 150\,\mathrm{km}$ from the center of mass and
extrapolated to null infinity as also described in
Ref.~\cite{Kyutoku:2014yba}.
\end{document}